\documentclass[runningheads]{svmult}
\usepackage{makeidx}   
\usepackage{graphicx}  
\usepackage{subeqnar}  
\usepackage{multicol}  
\usepackage{cropmark} 
\usepackage{physprbb}  
\def\lsim{\, \lower2truept\hbox{${<
\atop\hbox{\raise4truept\hbox{$\sim$}}}$}\,}

\def\gsim{\, \lower2truept\hbox{${>
\atop\hbox{\raise4truept\hbox{$\sim$}}}$}\,}

\begin{document}

\title*{The Intriguing Distribution of Dark Matter\\ in Galaxies}

\titlerunning{Dark Matter Distribution in Galaxies }

\author{Paolo Salucci\inst{1}
\and Annamaria Borriello\inst{1}}

\authorrunning{Salucci  \& Borriello}

\institute{(1) International School for Advanced Studies SISSA-ISAS -- Trieste, I}

\maketitle

\begin{abstract}

We review  the most recent evidence for the amazing properties of the density distribution of the
dark matter around spiral galaxies. Their rotation curves, coadded according to the galaxy luminosity, 
conform  to an Universal profile which can be represented as the sum of an exponential thin disk 
term plus a spherical halo term with a flat density core. From dwarfs to giants, these halos feature
a constant density region of size $r_0$ and core density $\rho_0$ related by 
$\rho_0= 4.5 \times 10^{-2} (r _0/{\rm kpc})^{-2/3} {\rm M}_{\odot} {\rm pc}^{-3}$.  
At the highest masses $\rho_0$ decreases exponentially  with $r_0$, revealing a lack of 
objects with disk masses $> 10^{11}M_{\odot}$ and central densities 
$> 1.5 \times 10^{-2}\ (r_0/{\rm kpc})^{-3} M_{\odot} {\rm pc}^{-3}$ 
implying a {\it maximum} mass of $\approx 2 \times 10^{12} M_{\odot}$ for a  dark  
halo hosting a stellar disk. The fine structure of dark matter halos is obtained from the
kinematics of  a number of suitable low--luminosity disk galaxies. The halo circular 
velocity  increases linearly with radius out to the edge of the stellar disk,
implying a constant dark halo density over the entire disk region.  The properties of halos 
around normal spirals provide substantial evidence of a discrepancy between the mass 
distributions predicted in the Cold Dark Matter scenario and those actually detected 
around galaxies.
\end{abstract}

\begin{figure}
\begin{center}
\vspace{1truecm}
\includegraphics[width=.5\textwidth]{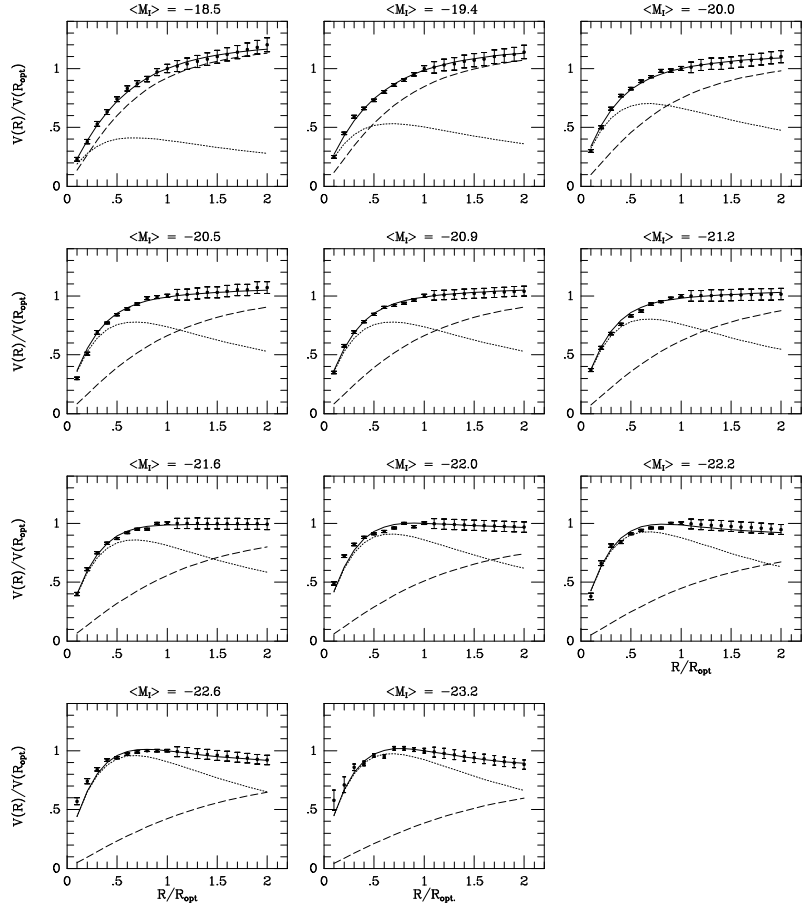}
\end{center}
\caption[]{Synthetic rotation curves (filled circles with error bars) and URC (solid line) 
with the separate dark/luminous
contributions (dotted line: disks; dashed line: halos).}
\end{figure}

\section{Introduction}

Rotation curves (RC's) of disk galaxies are the best probe for dark matter (DM) 
on galactic scale. Notwithstanding the impressive amount of knowledge gathered in 
the past 20 years, only very  recently we start to shed light to crucial aspects 
of the mass {\it distribution} of dark halos, including their radial density profile, 
and its claimed universality.
On a cosmological side, high--resolution  N--body simulations have shown 
that cold dark matter (CDM) halos achieve a specific equilibrium density 
profile [13 hereafter NFW, 5, 8, 12, 9] characterized by one free
parameter, e.g. the halo mass. In the inner region the DM halos density
profiles show  some scatter around an average profile which is characterized
by a power--law  cusp $\rho \sim r^{-\gamma} $, with $\gamma =1-1.5$ [13, 12, 
2].  In detail, the DM density profile is:
\begin{equation}  
\rho_{\rm NFW}(r) = \frac{\rho_s}{(r/r_s)(1+r/r_s)^2} 
\end{equation}
where $r_s$ and $\rho_s$  are respectively the characteristic inner radius and density. 
Let us define $r_{\rm vir}$ as the radius within which the mean density is
$\Delta_{\rm vir}$ times the mean universal density $\rho_m$ at the halo formation  redshift, 
and the associated virial mass $M_{\rm vir}$ and velocity $V_{\rm vir} \equiv G M_{\rm
vir} / r_{\rm vir}$. Hereafter we assume the $\Lambda$CDM scenario,  with $\Omega_m = 0.3$,
$\Omega_{\Lambda} =0.7$ and $h=0.75$, so that  $\Delta_{\rm vir} \simeq 340$ at $z \simeq 0$.
By assuming the concentration parameter as $c \equiv r_{\rm vir}/r_s$  the halo circular 
velocity $V_{\rm NFW}(r)$ takes the form [2]:
\begin{equation}
V_{\rm NFW}^2(r)= V_{vir}^2 \frac{c}{A(c)} \frac {A(x)}{x}
\end{equation}
where $x \equiv r/r_s$ and $A(x) \equiv \ln (1+x) - x/(1+x)$.
As the relation between $V_{ vir}$ and $r_{ vir}$ is fully specified by the background cosmology, 
the independent parameters characterizing the model reduce from three to two
($c$ and $r_s$). Let us stress that a high density $\Omega_m=1$ model, with a
concentration parameter $c>12$,  is definitely unable to account for the
observed galaxy kinematics [11]. 

So far, due to the limited number of suitable RC's and to the serious
uncertainties  in deriving  the actual  amount of luminous matter inside  the
inner regions of spirals, it has been difficult to investigate  the  internal
structure of dark halos. These difficulties have been overcome by means of:\\
\noindent {\it i)} a specific investigation of the Universal Rotation Curve [16], 
built by coadding $~ 1000$ RC's, in which we adopt a  general halo
mass distribution: 
\begin{equation} 
V_{h,URC}^2(x)= V^2_{opt}\ (1-\beta)\ (1+a^2)\ {x^2 \over (x^2+a^2)}
\end{equation} 
with $x \equiv r/r_{opt}$, $a$ the halo core radius in units of
$r_{opt}$   and $\beta \equiv (V_{d,URC}(r_{opt})/V_{opt})^2$.
It is important to remark that, out to $r_{opt}$, this  mass  model is {\it
neutral}  with respect to  the halo profile.  Indeed, by varying
$\beta$ and $a$, we  can  efficiently reproduce the maximum--disk,  the
solid--body, the  no--halo, the all--halo, the CDM and the core-less--halo
models.  For instance, CDM halos with concentration parameter $c=5$ and
$r_s=r_{opt}$  are well fit by (3) with $a \simeq 0.33$\\   
\noindent {\it ii)} a number of suitably selected individual RC's [1], whose mass
decomposition has been made adopting the cored Burkert--Borriello--Salucci
(BBS) halo profile (see below).

\section{Dark Matter Properties from the Universal Rotation Curve}

The observational framework is the following: {\it a}) the mass in spirals 
is distributed  according to the Inner Baryon 
Dominance (IBD) regime [16]: there is  a characteristic transition radius 
$r_{IBD} \simeq 2  r_d (V_{opt}/220 \ {\rm km/s})^{1.2}$  ($r_d$ is the disk
scale--length and $V_{opt} \equiv V(r_{opt})$) according which,  for $r \leq
r_{IBD}$, the luminous matter totally  accounts for the gravitating mass, 
whereas,  for $r > r_{IBD}$,  the dark matter shows dynamically up and {\it
rapidly}  becomes the dominant   component [20, 18, 1].   Then, although dark
halo might extend down to the galaxy centers,  it is only for $r > r_{IBD}$
that they give a non--negligible contribution to the circular velocity.  {\it
b}) DM is distributed in a different way with respect to any of the various
baryonic components [16, 6],  and {\it c}) HI contribution to the circular
velocity at $r < r_{opt}$, is negligible [e.g. 17].

\subsection{Mass modeling}

Persic, Salucci and Stel [16] have derived  from  $\sim 15000$ velocity
measurements of $\sim 1000$ RC's the synthetic rotation velocities of spirals
$V_{syn} ({r \over {r_{opt}} }; {L_I\over{L_*}})$,  sorted by  luminosity 
(Fig. 1, with $L_I$ the  $I$--band luminosity ($L_I/L_*=10^{-(M_I+21.9)/5}$).
Remarkably, individual RC's  have a negligible variance with respect to their
corresponding synthetic curves: spirals sweep a very narrow locus in the
RC-~profile/amplitude/luminosity space. In addition,  kinematical   properties
of spirals do significantly change with galaxy  luminosity [e.g. 16], then it
is  natural  to relate their mass distribution with this quantity. The whole
set of synthetic  RC's define the Universal Rotation Curve (URC), composed 
by the sum of two terms: {\it a}) an exponential thin disk with circular
velocity (see [16]): 
\begin{equation}
V^2_{d,URC}(x)=1.28~\beta\ V^2_{opt}~ x^2~(I_0K_0-I_1K_1)|_{1.6x}
\end{equation}
and a spherical halo, whose velocity contribution is given by (3).
At high luminosities,  the contribution from a bulge component 
has also been  considered.

The data (i.e. the synthetic curves $V_{syn}$) select the actual model out of this
family, by setting  $V_{URC}^2(x)=V^2_{h,URC}(x,\beta,a)+ V^2_{d, URC}(x,\beta) $ 
with $a$ and $\beta$ as free parameters. An extremely good fit occurs for   
$a \simeq 1.5 (L_I/L_*)$ [16] or, equivalently, for  $a =a(\beta)$
and $ \beta=\beta (\log V_{opt})$ as plotted in Fig. 2. With these values the
URC  reproduces the data $V_{syn}(r)$ up to their {\it rms} (i.e.  within
$2\%$).  Moreover,  at fixed luminosity the $\sigma$ fitting uncertainties
in  $a$ and $\beta$ are lesser than   20\%. The emerging  picture is: {\it
i)} smaller objects have more fractional amount of dark matter
(inside $r_{\rm opt}$:  $M_{\ast}/M_{\rm vir} \simeq 0.2\ (M_{\ast}/2 \times
10^{11} M_{\odot})^{0.75}$ [20]), {\it ii)} dark mass increses
with radius much more that linearly.

\begin{figure}[t]
\begin{center}
\vspace{-3.8truecm}
\includegraphics[width=.7\textwidth]{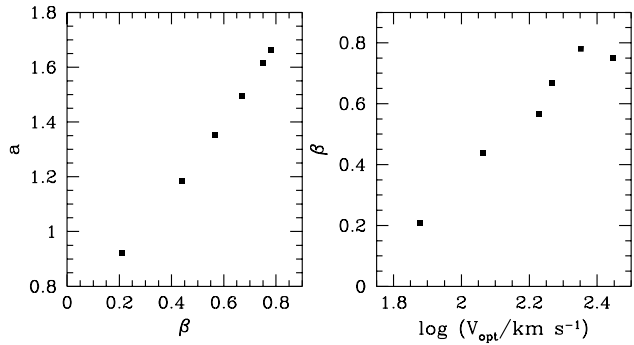}
\end{center}
\caption[]{$a$ {\it vs.} $\beta$ and  $\beta$ {\it vs.} $V_{opt}$.}
\end{figure}

\subsection{Halo Density Profiles}

The above  evidence calls for a quite specific DM density profile; 
we adopt the BBS halo mass distribution [3, 4, 1]:
\begin{equation}
\rho_{\rm BBS}(r) = \frac{\rho_0\ r^3_0}{(r+r_0)(r^2+r_0^2)}
\end{equation}
where $\rho_0$ and $r_0$ are free parameters which represent the central 
DM density and the core radius. Of course, for $r_0 \ll r_d$, we recover a
cuspy profile. Within spherical symmetry, the mass distribution   is given by:
\begin{equation}
M_{\rm BBS}(r) = 4\ M_0\ \{ \ln (1 + r/r_0)  -\arctan (r/r_0) + 0.5  \ln  [1
+(r/r_0)^2]\}
\end{equation}
\begin{equation}
M_0 \simeq 1.6\ \rho_0\ r_0^3  
\end{equation}
with $M_0$ the dark  mass within the core. 
The halo contribution to the circular velocity is then: $V^2_{\rm
BBS}(r) =G\ M_{\rm BBS}(r)/ r$. Although the dark matter ``core" parameters
$r_0$, $\rho_0$ and $M_0$ are in principle independent,  the observations
reveal a quite strong  correlation among them [e.g. 19]. Then, dark halos  may
be an 1--parameter family, completely specified by e.g. their core mass $M_0$.
When we test the disk+BBS  velocities with $\rho_0$ and $r_0$ left as free
parameters, we find that, at any luminosity  and  out to $\sim 6\ r_d$, the
model  is indistinguishable from data  (i.e. $V_{syn}(r)$). 
\begin{figure}[t]
\vskip -1.4truecm
\hskip 0.1truecm
\includegraphics[width=1.2\textwidth]{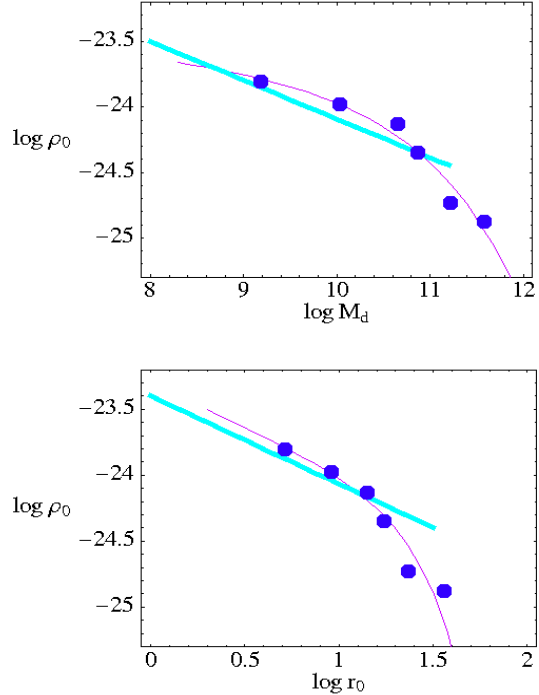}
\vskip -8.7truecm
\caption[]{ {\it up)} Central halo density $\rho_0$  (in ${\rm g/cm}^3$)
{\it vs.}  disk mass (in solar units) for normal spirals ({\it filled
circles}); {\it bottom)} central
density {\it vs.} core radii (in kpc) for normal  spirals. 
The  straight lines are from [3],
whereas the curved lines are the best fits used in \S 4.}
\end{figure}
More specifically,
we reproduce   the synthetic rotation curves at the level of their {\it rms}.
The values of  $r_0$ and $\rho_0$ derived in this way   agree with the
extrapolation at high masses   of the scaling law $\rho \propto r_0^{-2/3} $
[3] established for objects with much smaller core radii $r_0$ and  stellar
masses (see Fig. 3). Let us notice that the core radii are pretty large ($r_0
\gg r_d$): ever-rising halo RC's cannot be excluded by the data. Moreover,
spirals lie on the extrapolation of the disk--mass  {\it vs.} central halo density 
relationship $\rho_0 \propto M_d^{-1/3}$  found for dwarf galaxies [3],  to
indicate that the densest halos harbor the least massive disks (see Fig. 3).

The curvature in $\rho_0$ {\it vs.} $r_0$ at the highest
masses/lowest densities can be linked to  the existence of an {\it upper mass
limit} in $M_{\rm vir}$ which is evident by the sudden decline of
the baryonic mass function of  disk galaxies at $M_d^{max}= 2\times
10^{11}M_\odot$ [20]. In fact, such a limit implies a maximum halo mass
of  $M_{\rm vir}^{max} \sim \Omega_0/\Omega_{b}\ M_d^{max}$. Then, for 
(6) and (7),  $M_{\rm vir} = \eta\ M_0$,  with $\eta \simeq 12$  for
($\Omega_0$, $\Omega_b$ ,$z$) = (0.3, 0.03, 3), and the limiting halo mass
implies a lack of objects with $\rho_0> 4\times 10^{-25}\ {\rm
g/cm}^{3}$ and $r_0>30 \ {\rm kpc}$, as is evident in Fig. 3.  On the other
side,  the observed  deficit of objects with  $M_d \sim M_d^{max} $ and
$\rho_0> 4\times 10^{-25}~{\rm g/cm}^{3}$, suggests that, at this mass scale,
the total--to--baryonic density ratio nears the cosmological value
$\Omega_0/\Omega_{b}\simeq 10$.

\subsection{Testing CDM with the URC}

The BBS  density profile reproduces in synthetic RC's the DM halo contributions, at least
out to two optical radii.  This is in contradiction with CDM halo properties according to 
which  the velocity dispersion $\sigma$ of the dark matter particles decreases
towards the center to reach $\sigma \rightarrow 0$ for $r \rightarrow 0$. 
Dark halos therefore, are not kinematically cold structures,  but
 ``warm" regions of  sizes $ r_0 \propto \rho_0^{-1.5}$  which, by the way, 
turn up quite large: $r_0 \sim 4-7\ r_d$. Then, the boundary of the core region  
is well beyond the region  of the stellar disk and there is not evidence that dark halos 
converge to a $\rho \sim r^{-2}$ (or a steeper) regime, as dictated by CDM
predictions.

\section{Dark Matter Properties from Individual Rotation Curves }

Although deriving halo densities from individual RC's is certainly complicated,   
the belief according to which  one always gets ambiguous  halo  mass modeling 
[e.g. 22] is incorrect. In fact, this is true only for rotation curves of low spatial resolution, 
i.e. with less than $\sim3$ measures per exponential disk length--scale
occurring in  most HI RC's.  In this case, since the parameters of the galaxy structure 
are very sensitive to the {\it shape} of the rotation curve in the region $0<r<r_d$,
there are no sufficient data to constrain  models.

In the case of  high--quality {\it optical} RC's  tens of independent 
measurements in  the critical region make possible to infer the halo mass distribution. 
Moreover, since the dark component can be better traced when the disk contributes 
to the dynamics in a modest way, a convenient strategy leads  to investigate DM--dominated 
objects, like dwarf and low surface brightness (LSB) galaxies. It is well known that for 
the latter   [e.g. 7, 11, 3, 4, 9, 10, 21] the results are far from being definitive in that they are
{\it 1)} affected by a quite  low spatial resolution and {\it 2)} uncertain, due to the limited 
amount of available kinematical data [e.g. 23].

Since most of the properties of cosmological halos are claimed universal,  an
useful strategy is to investigate  a number of high--quality {\it
optical} rotation curves  of {\it low luminosity } late--type spirals, with
$I$--band absolute magnitudes $-21.4<M_I<-20.0$  and  $100 <V_{opt}< 170$ km
s$^{-1}$. Objects in this luminosity/velocity range are DM dominated [e.g. 20]
 but their RC's, measured at an angular resolution of $2^{\prime \prime}$, 
have an excellent  spatial resolution of  $\sim 100 (D/10 \ {\rm Mpc})$  pc and $n_{data}
\sim r_{opt}/w$ independent measurements.  For nearby galaxies: $ w<< r_{d}$
and $n_{data}>25$.   Moreover, we select RC's of bulge--less systems, so that
the stellar disk is the only baryonic component for $r \lsim r_d$.

In detail, we extract the best 9 rotation curves, from the `excellent' subsample  of $80$ rotation curves of [15], 
which are all suitable for an accurate mass modeling. In fact,  these RC's trace properly  
the gravitational potential in that: {\it 1)}  data extend at least out to the optical radius, {\it 2)} they
are smooth and  symmetric, {\it 3)} they have small {\it rms}, {\it 4)} they
have high  spatial resolution and a homogeneous radial data coverage, i.e.
about  $30-100$ data points homogeneously distributed with radius and between
the  two arms.  
The 9 extracted galaxies are of low luminosity  
($5 \times 10^9 L_{\odot} <L_I< 2 \times  10^{10} L_{\odot}$; $100<V_{opt}< 170$ km s$^{-1}$)
and  their $I$--band surface luminosity profiles are (almost) perfect radial exponential. 
These two last criteria, not indispensable to  perform the {\it mass} decomposition, help inferring 
the  dark halo {\it density}  distribution.  Each RC has $7-15$
velocity points inside  $r_{opt}$, each one being the average of $2-6$ independent data. 
The RC's  spatial resolution is better than $1/20\ r_{opt}$, the velocity {\it
rms}  is about $3\%$ and  the RC's logarithmic derivative is generally known
within  about 0.05.    
 
\begin{figure}[t] 
\begin{center}
\includegraphics[width=.5\textwidth]{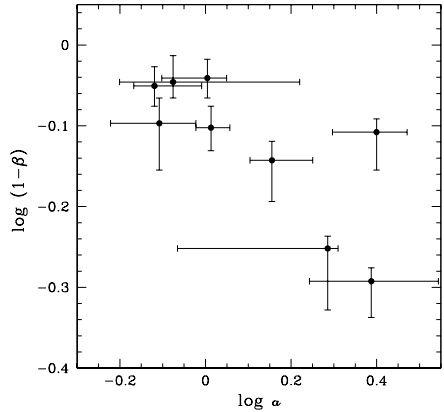}
\end{center}
\vspace {-0.2truecm}
\caption[]{Halo parameters ($a$ is in units of $r_{opt}$) for the individual
RC's. Notice that for this sample we can derive in model--independent way $a
\gsim 1$, in disagreement with CDM predictions.} 
\end{figure}

\subsection{ Halo Density Profiles}
  
We model the mass distribution as the sum of two components: a stellar disk 
and a spherical dark halo. By assuming centrifugal equilibrium under the 
action of the gravitational potential, the observed circular velocity can be 
split into the two components:  $V^2(r)=V^2_d(r)+V^2_h(r)$.  
By selection, the objects are bulge--less and stars are distributed like an
exponential thin disk. Light traces the mass via an assumed radially constant
mass--to--light ratio. 

We neglect the gas contribution
$V_{gas}(r)$ since   in normal spirals it is usually modest within the optical
region [17,  Fig. 4.13]: $ \beta_{gas}\equiv (V^2_{\rm gas}/V^2)_{r_{opt}}
\sim 0.1$.   Furthermore, high resolution HI observations show that in  low
luminosity spirals:  $V_{gas}(r) \simeq 0$ for $r < r_d$ and  $V_{gas}(r)
\simeq (20 \pm 5) (r-r_d) / 2 r_d $ for   $r_d \leq r \leq 3r_d $.  Thus, in
the optical region:  {\it i)} $V_{gas}^2(r)<<V^2(r)$ and {\it ii)} 
$d[V^2(r)-V^2_{gas}(r)]/dr \gsim 0$. This last condition implies that 
by  including $V_{gas}$ the halo velocity profiles  
would result {\it steeper} and then the
core radius in the halo density even {\it larger}. Incidentally, this is not
the case for dwarfs and LSB´s: most  of their kinematics is affected by the HI
disk gravitational pull in such  a way that neglecting it could bias the
determination of the DM density.  The circular velocity profile of the disk
is given by (4) and the DM halo will have the form given by (3).  Since we
normalize (at  $r_{opt}$) the velocity model $(V_d^2+V^2_h)^{1/2}$ to the
observed rotation  speed $V_{opt}$, $\beta$ enters explicitly in the halo
velocity model and this reduces the free parameters  of the mass model to
two.   

\begin{figure}[t]
\vspace{-1.6truecm}
\begin{center}
\includegraphics[width=12.5truecm,height=17truecm]{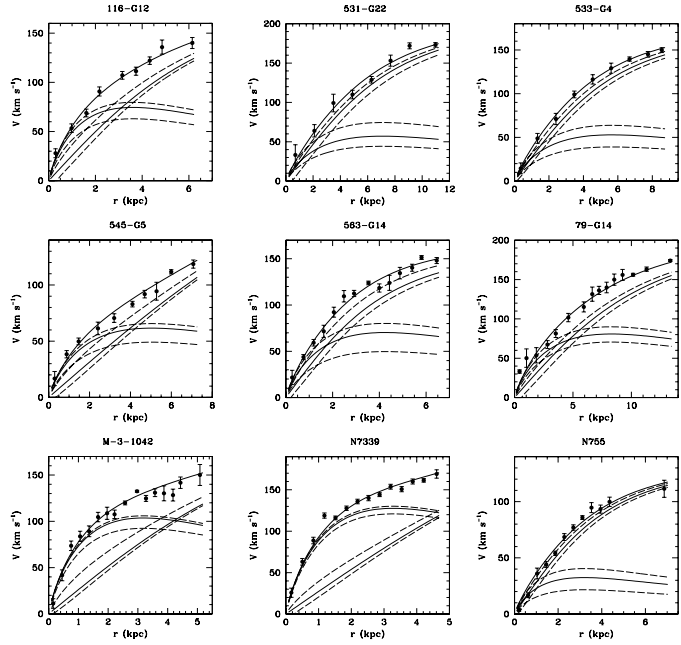}
\end{center}
\vspace{-6.7truecm}
\caption[]{ BBS fits ({\it thick solid line}) to the RC's ({\it points with
errorbars}).  Thin solid lines represent the   disk and halo contributions.
Notice the steep halo velocity profiles. The maximum and minimum  disk
solutions  ({\it dashed lines}) provide the theoretical uncertainties.}
\end{figure}

For each galaxy, we determine the values of the parameters $\beta $ and $a$ 
by means of a $\chi ^2$--minimization fit to the observed rotation curves: 
\begin{equation} 
V^2_{model}(r; \beta , a) = V^2_d (r; \beta) + V^2_h (r; \beta , a) 
\end{equation} 
A central role in discriminating among the different mass decompositions is  
played by the derivative of the velocity field $dV/dr$. It has been shown  
[e.g. 14] that by taking into   account the logarithmic gradient of the
circular velocity field defined as:  $\nabla (r)\equiv \frac{d \log V(r)}{d \log r} $  
one can retrieve the  crucial  information  stored in the shape of
the rotation curve.   Then,  we set the    $\chi^2$'s  as the sum of those
evaluated on velocities and  on logarithmic  gradients:  
$ \chi^2_V =\sum^{n_V}_{i=1}\frac{V_i-V_{model}(r_i; \beta,a)}  {\delta V_i}$ 
and  $\chi^2_{\nabla} = \sum^{n_{\nabla}}_{i=1}\frac{\nabla(r_i)-\nabla_{model}
(r_i; \beta,a)}{\delta \nabla_i}$,
with  $\nabla_{model} (r_i, \beta,a)$ given  from the above equations.
The parameters  of the mass models are finally  obtained   by minimizing the
quantity $\chi^2_{tot} \equiv \chi^2_V+ \chi^2_{\nabla}$.  

\begin{figure}[t]
\vspace{-1.6truecm} 
\begin{center}
\includegraphics[width=12.5truecm, height=17truecm]{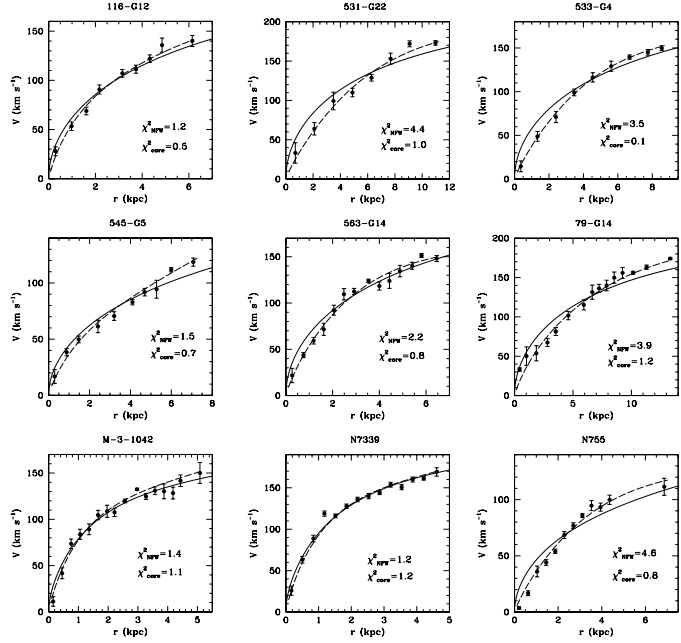} 
\end{center}
\vspace{-6.5truecm} 
\caption[]{NFW best--fits {\it solid lines} of the rotation curves  {\it
(filled circles)} compared with   the BBS fits {\it (dashed lines)}. The  
$\chi^2$ values are also indicated.} 
\end{figure} 
 
The best--fit models parameters are shown in Fig. 4. The
disk--contribution $\beta $ and the  halo core radius $a$ span a range from
0.1 to 0.5 and from 0.8 to 2.5, respectively. They are pretty well constrained in
a small and continuous region of the ($a$, $\beta$)  space.  The derived mass
models are shown in Fig. 5, alongside with the separate disk and halo
contributions.  We also get a ``lowest" and a ``highest" halo velocity curve
(dashed lines in figure) by subtracting from $V(r)$ the maximum and the
minimum disk contributions $V_d(r)$ obtained by substituting in (4) the
parameter $\beta$ with $\beta_{best}+\delta\beta$ and
$\beta_{best}-\delta\beta$, respectively. It is obvious that the halo curves
are steadily increasing, almost linearly,  out to the last data point. In each
object the uniqueness of the resulting halo velocity model can be realized by
the fact that the maximum--disk and minimum--disk models almost coincide.
Remarkably, we find that the size of the halo density core is always greater
than the disk characteristic scale--length $r_d$ and it can extend beyond the
disk edge (and the region investigated).

\subsection{Testing CDM}

In Fig. 5  we show how  the halo velocity profiles of the nine galaxies rise 
almost linearly with radius, at least out to the disk edge:
\begin{equation}
V_h(r) \propto r \ \ \ \ \ \ \ \ \ 0.05\ r_{opt} \lsim r\ \lsim r_{opt}
\end{equation}
The halo density profile has a well defined (core) radius within which
the density is approximately constant. This is inconsistent with the
singular halo density distribution emerging in the Cold Dark Matter (CDM)
halo formation scenario. More precisely, since the CDM
halos are, at small radii, likely more cuspy than the NFW profile:
$\rho_{CDM}\propto r^{-1.5}$ [e.g. 14], the steepest CDM halo
velocity profile $V_h(r) \propto r^{1/4}$ results too shallow with
respect to observations.
Although the mass models of (3) converge to a distribution  with an inner
core rather than with a central spike, it is worth, given the importance of such result,  
checking in a direct way the (in)compatibility of the CDM  models with galaxy  kinematics.  

So, we assume the NFW functional form for the halo density  given by (1), 
leaving $c$ and $r_s$ as free independent parameters,
although N--body simulations and semi-analytic investigations indicate  that
they correlate. This choice  to increase the chance of a good  fit. We also imposed
to the object under study a conservative halo mass upper limit of
$2 \times 10^{12} M_\odot$.
The fits to the data are shown in  Fig. 6, together with the BBS fits: for
seven out of  nine   objects the NFW models are unacceptably worse than the
BBS solutions.   Moreover,  in all objects,  the CDM  virial mass is too high:
$M_{vir} \sim 2 \times 10^{12} M_{\odot}$  and  the resulting  disk 
mass--to--light ratio too low ($\lsim 10^{-1}$ in the $I$-band).

\section{The Intriguing Evidence from Dark Matter Halos}

The dark halos around spirals emerge as an one--parameter family;
the order parameter (either the central density or the core
radius)  correlates with the luminous mass. However, we do not know how it is
related to the global structural properties of the dark halo, like the 
virial radius or the virial mass, unless we extrapolate out over the BBS
profile.  That is because the halo RC, out to the outermost data, is completely
determined by physical parameters, the central core density and the core
radius,  which have little counterpart in the gravitational
instability/hierarchical clustering picture.   
\begin{figure}[t] 
\begin{center}
\includegraphics[width=.9\textwidth]{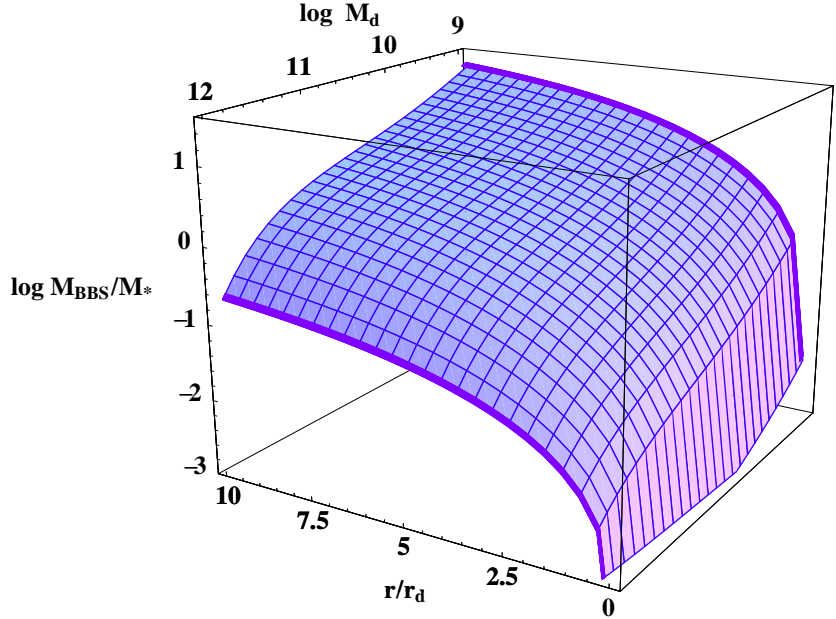}
\end{center}
\vspace {-0.2truecm}
\caption[]{The dark--to--luminous mass ratio as function of the normalized 
radius and the total disk mass.} 
\end{figure}

Caveat the above extrapolation, the location of spiral galaxies in the virtual
space of virial mass/halo ``central" density/stellar mass, that, on
theorethical basis, should be roughly 3-D random determined by several
different and non--linear physical processes, is remarkably found to degenerate
and to lie on a curve.  Indeed, in Fig.7 we show the dark--to--luminous mass
ratio as function of the normalized  radius and the total disk mass. The
surface has been obtained by adopting the correlations between the halo and
the disk parameters we found in our previous works (see Fig.3 [19]):
\begin{eqnarray}
\log r_0 =   9.10 +  0.28\ \log \rho_0  -3.49 \times 10^{10} \ \rho_0^{0.43} \\ 
\log \rho_0  = -23.0 -0.077\ \log M_d - 9.98 \times 10^{-6}\  M_d^{0.43}
\end{eqnarray} 
and:
\begin{equation}
\log r_d  = 4.96 - 1.17\ \log M_d + 0.070\ (\log M_d)^2 
\end{equation}
from data in  [16].   The  dark--to--luminous mass ratio at fixed ratio increases as  the total disk mass
decreases;  for example at $r=10\ r_d$ it raises from $20\%$ for massive disks ($M_d=10^{12} M_{\odot}$) 
to $220\%$  for  smaller disks ($M_d=10^9 M_{\odot}$).                                                                                                     
  
Two conclusive statements can be drawn:   dark matter halos
have an inner  constant--density region, whose size exceeds the stellar 
disk length--scale. Second, there is no evidence that dark halos converge, at
large radii, to a $\rho \sim r^{-2}$ (or steeper) profile. 

The existence of a region of  ``constant"  density   $\rho_0 \simeq
{\pi \over 24} \ {M_{\rm vir} \over r_0^3}$ and size $r_0$  is hardly
explained  within current theories of galaxy formation.  Moreover, 
the evidence of a smooth halo profile is growing more an more 
in recent  literature [e.g. 26, 27, 28, 29] and a number of different 
solutions have been proposed to solve this problem [e.g. 30, 31, 32, 33].
Let us stress, however, that we should  incorporate all the intriguing 
halo properties described in this review.

\end{document}